\documentclass[conference]{IEEEtran}
\IEEEoverridecommandlockouts
\usepackage{cite}
\usepackage{color,soul}
\usepackage{amsmath,amssymb,amsfonts}
\usepackage{algorithmic}
\usepackage{multicol}
\usepackage{graphicx}
\usepackage{subcaption}
\usepackage{textcomp}
\usepackage{xcolor}
\usepackage{hyperref}
\def\BibTeX{{\rm B\kern-.05em{\sc i\kern-.025em b}\kern-.08em
    T\kern-.1667em\lower.7ex\hbox{E}\kern-.125emX}}
\begin{document}

\title{fybrrChat: A Distributed Chat Application for Secure P2P Messaging\\
}

\author{\IEEEauthorblockN{Debajyoti Halder\IEEEauthorrefmark{2},
Saksham Bhushan\IEEEauthorrefmark{2},
Gundu Shreya\IEEEauthorrefmark{2}, and
Prashant Kumar\IEEEauthorrefmark{3}}
\IEEEauthorblockA{\IEEEauthorrefmark{2}Department of Electrical Engineering and Computer Science,
Indian Institute of Technology Bhilai, India}
\IEEEauthorblockA{\IEEEauthorrefmark{3}Software Engineer, Atonarp Inc., India}
}


\maketitle

\begin{abstract}
The demand for connecting with others and sharing messages and multimedia files is ever-growing but people's lack of trust in a centralized system makes the system unusable for extremely critical and sensitive use cases such as military and government operations. Therefore, there is an urgent need for a system that ensures impenetrable security embedded into the methodology used for communication between peers. Therefore, we propose a decentralized secure chat application based on trustless network architecture, which avoids any central authority or organisation from exploiting users by accessing their private data or manipulating the user-end policies. We also implement end-to-end encryption for the secure transfer of messages between peers. To enforce a critical policy change, in the absence of a central authority, we provide a Consensus mechanism where peers in the network vote for the particular policy change within a private or public swarm.  
\end{abstract}

\begin{IEEEkeywords}
Peer-to-Peer, Distributed File Storage, Distributed Hash Table, Distributed Message Queue
\end{IEEEkeywords}

\vspace{-0.6em}
\section{Introduction}
The growing demand for connecting with each other across the world has proved to be a boon to the growth of social media platforms. But when it comes to ensuring the privacy and security of the platform, the control is in hands of few monopolies. Some claim to provide a secure medium of communication but their exploitation of users and misusing users' data with their centralized control over the policies has led to various controversies. Lack of trust towards these organisations have made people skeptical towards using these applications for extremely secure communication use-cases. Thus, a system with no central organisation in control and a \emph{trustless} network \cite{trustless} is required. A trustless network is a network which does not requires the participants (users) to trust any central authority. These nodes themselves, as a group, can be trusted to make a decision which aligns with the interests of all. This is accompanied by a distributed network, where peers are connected to each other to form a mesh-like network called Swarm. This distributed network of connected peers is often attributed as the internet of tomorrow \cite{dist}.
The total number of monthly active users on the top 3 chat applications surpassed 4 billion users in July 2019. As of Oct 2020, WhatsApp alone has 2 billion monthly active users sending around 100 billion messages per day \cite{stats}. With such vast number of people relying on chat applications, making these applications highly secure becomes a task of paramount importance.

To make a chat application with trustless network, it is required that the whole system is based on peer-to-peer protocols and no server to store the messages centrally. It becomes extremely difficult to reliably transfer a message to the destination without a central server. This problem becomes critical when the peers are not online to receive the message instantaneously. Centralized systems depend on servers with message queues which make sure that the message is surely delivered to the destination. Therefore, to solve this, we also incorporate message queues but in a distributed manner.

For extremely critical and secure messaging use cases, such as military or government operations, the security threat is not just from the central organization managing/owning the application but also from other peers in the network who can try to penetrate the communication channel between two users. Thus, we address this critical issue by providing an option for private swarm. Private Swarm is a private network comprised of only authorised peers of an organisation. For example, government can use a private swarm and authorise devices only of other government employees involved in the particular decision making. This will not just enhance the security of the system but also make the application ubiquitous in all use cases, ranging from normal social interaction to highly sensitive communications such as in military.

We propose a distributed chat application which can be used to share files and messages securely between users. We propose a trustless mechanism which avoids any control over the content shared or users' data. We also propose a consensus mechanism which enables the users of the network to make decisions as a group to benefit everyone in the network. We also enhance the security of a network by proposing a private swarm functionality, which allows only authorised devices to be a part of the network making it a highly secure isolated network. We also propose distributed messaging queues which use Distributed File Storage and Distributed Hash Tables (DHT) to securely manage undelivered messages.

\section{Related Works}

The most common instant messaging applications (WhatsApp \cite{wp}, Telegram \cite{telegram}, Messenger \cite{messenger}) that we use daily are based on client-server architecture, which is centralized and stores users' data in the cloud. There have been a few pieces of research on some alternative solutions that propose an approach that are not completely centralized. The proposed solution in \cite{pubsub1, pubsub2} is a P2P topic publish/subscribe based algorithm that floods (publish) the network with the sent message on a particular topic, all the nodes that are subscribed to the topic will receive the message. The downside of this approach is that the message is flooded in the network which is a wastage of bandwidth. If sent through a socket server to decrease bandwidth wastage, the solution does not remain P2P anymore. We solve this problem using a direct P2P data channel to send the messages. This decreases the latency in message transmission. Also, the solutions are unable to send undelivered messages to the recipient since they do not have a method of storing the undelivered message. To implement this feature they need to use a cloud service which misses the whole purpose of this approach. We solve this using a Distributed Message Queue (DMQ) that stores the undelivered messages until delivered. The complete approach has been explained in Section \ref{solution}.

\section{Proposed Solution}
\label{solution}

The application can be used for instant messaging between two or more users' at once. To connect to the distributed network the user can run the chat application as a web-based application or a mobile application.

Using this model, the users can chat or transfer files to each other without the requirement of a central entity that is responsible for relaying or storing the message. The users can establish a peer-to-peer connection between each other upon initiating a chat when both the sender and receiver devices are online. This connection establishment is possible with the help of a signalling server which is responsible for discovering the other peer in the network. In case of the receiver device being offline at the time of message transmission by the sender, the message is stored in the distributed message queue and is located using a content addressing protocol by the receiver when the device is back online.
 
\subsection{Modules of the Architecture}
We have divided the various essential functionalities of the proposed distributed system for fybrrChat into the following modules for simpler explanation: 
    \subsubsection{\textbf{RTC Peer Connection}} When both the receiver and the sender are online, a Real Time Communication (RTC) Peer Connection is established between them. This is useful as it establishes a secure direct connection between the two devices using the Stream Control Transmission Protocol, which significantly reduces the delay in message transmission. In this case, the message circulation in the distributed network is bypassed which eventually reduces the load on the distributed network.
    \subsubsection{\textbf{End-to-End Encryption (E2EE)}} Every message sent out by a sender is encrypted with an end-to-end encryption protocol before leaving the senders' device. We use public-key cryptography where the message is encrypted by the receiver's public key and can only be decrypted with the private key of the receiver, which is confidential to the receiver.
    \subsubsection{\textbf{Content Addressing}} Since it can be difficult to track a message in a distributed system and also validating the authenticity of the message upon its reception is crucial, the message stored in the system is addressed with a content addressing protocol. The content addressing protocol uses hashing function to hash the content and instead of a location-based address, the system uses a content-based identifier for the message or file. The hash is itself the address of the content, which not only makes it easier for locating the content but also to validate the contents of the message. Also, since the content hash is the identifier, it makes the content immutable. Therefore, for each message and file, if changed even slightly, a new address will be assigned to it. In our system we use SHA-256 for content hashing.
    \subsubsection{\textbf{Content Pinning}} It is often the case that the sender after sending a message goes offline. In such a case the message sent by the user will have to stay in the network until it is delivered to the receiver. We solve this by storing the message in the distributed storage which comprises of the user devices currently in the network. Also, for the reliable retrieval of the message, redundancy across various nodes is required which is also achieved by content pinning. In content pinning, the message is broken into chunks and is circulated across some of the peers in the network for them to store the chunk until the receiver successfully receives the complete message. Few dedicated storage devices or user devices called as the bootstrap nodes can also be introduced in the network which will be responsible for faster storage and retrieval of the messages, thus, increasing the reliability of the system. Since, the message is encrypted and can only be decrypted with the private key of the receiver, it is impossible to read the contents of the message.
    \subsubsection{\textbf{Distributed Hash Table (DHT)}} The lookup of the contents in the distributed storage is done using DHT. The receiver of the message can look up in the DHT with the hash of the message for locating the message intended for it.
    \subsubsection{\textbf{Distributed Message Queue (DMQ)}} When the sender sends a message to a receiver who is offline, the message sent is stored in the message queue. The message queue stores the hash of the message in a distributed storage which is not controlled by any central entity. When the receiver comes online, it retrieves the available message hashes intended for it from the DMQ and then looks into the DHT for the actual encrypted message for each message hash.
    \subsubsection{\textbf{Consensus}} In our system, the lack of a centralized stake-holder can result in problems such as inconsistency in peers in the network, implementation of crucial policies and maintenance of the public/private swarm. Consensus will enable users as a group in the network to behave as the centralized stake-holder for consistency and maintenance of the network. 
    Using consensus, bootstrap nodes in the network can also be provisioned. The addition or removal of nodes in a private swarm can be done using consensus of the existing nodes in the swarm. Defining which peers are authorised for performing a certain task can also be determined using consensus.
    

\begin{figure}[h]
\centerline{\includegraphics[width=0.8\linewidth]{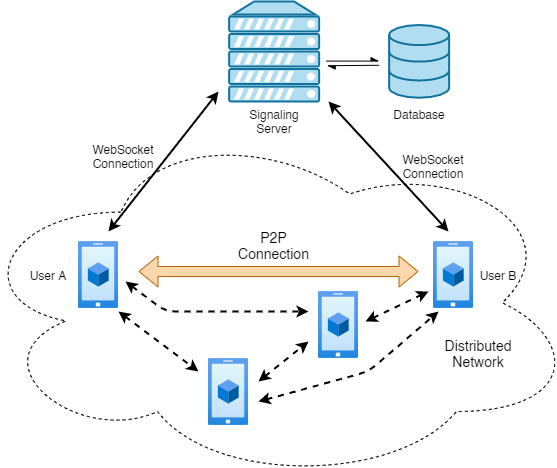}}
\caption{High-level architecture of the proposed system.}
\label{fig:arch}
\end{figure}

\begin{figure*}[h]
\centerline{\includegraphics[width=0.8\textwidth]{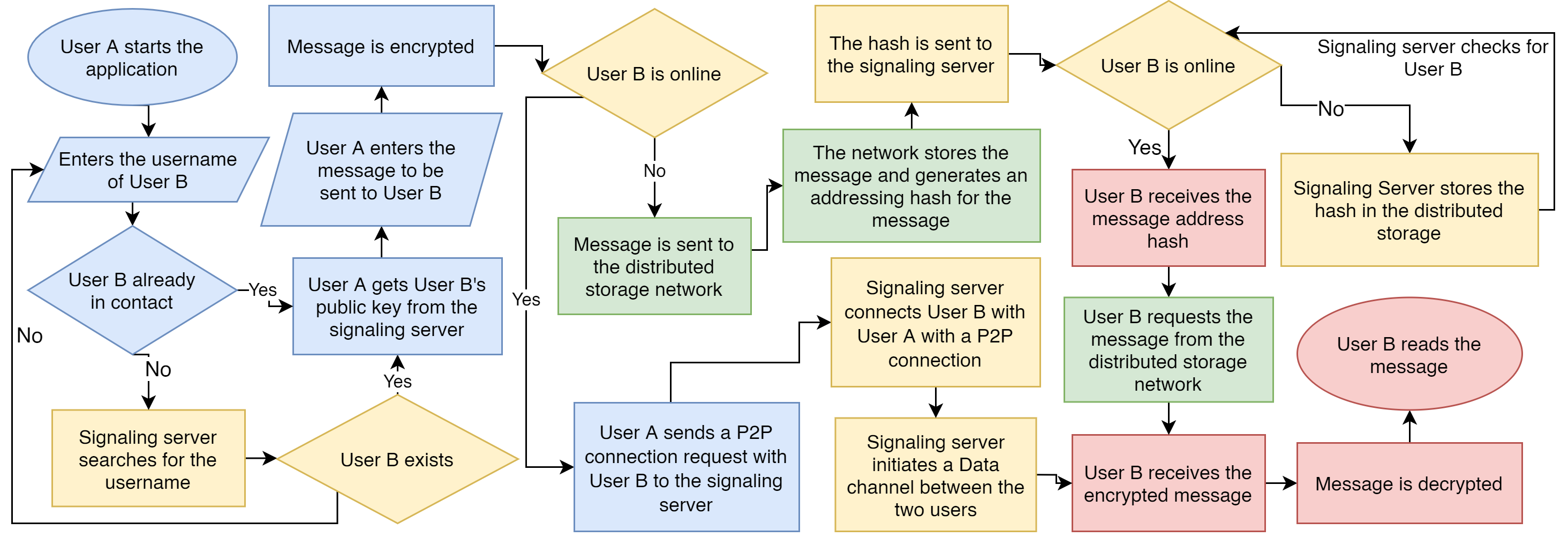}}
\caption{Flow of message through fybrrChat architecture.}
\label{fig:flow}
\end{figure*}

\subsection{Network Architecture}

The network architecture of the proposed peer-to-peer distributed chat application is given in Fig. \ref{fig:arch}. The architecture comprises of the following entities.



\subsubsection{\textbf{Peers}} Peers are the most essential components of the architecture as each user device in the network is a peer. They are the generator and consumer of the messages and files in the system. In the distributed storage system, peers are responsible for content pinning. The peers are connected to each other for maintaining the consistency of DHT using TCP or a WebSocket connection.

\subsubsection{\textbf{Signalling Server}} It is responsible for establishing a direct peer-to-peer RTC DataChannel between the users and it is also connected with the user authentication database. It initiates a handshake between peers to establish a RTC DataChannel using the Session Description Protocol. It is connected to the peers in the network using a WebSocket connection.

\subsubsection{\textbf{Database}} The database can be used for various purposes such as storing user authentication details through which user login can be authorised. It can also keep a note of active peers in the distributed network for faster connection establishment between users. For the current implementation, messaging queue is also implemented using this database.

\section{Performance Evaluation}

For fybrrChat prototype \cite{fybrrChat} we have implemented the following essential features required for messaging using some of the existing technologies mentioned in Table \ref{tab:1}. The flow of the message through the different modules of the architecture is illustrated in Fig. \ref{fig:flow}.

\begin{table}[h]
  \centering
    \caption{Technologies used for implementing features of fybrrChat.}
  \label{tab:1}
  \begin{tabular}{|p{3cm}|p{4.5cm}|}
  
  \hline 
    \textbf{Feature implemented} & \textbf{Technology used}\\ \hline
    \hline
    P2P RTC DataChannel & WebRTC \cite{webrtc}\\ \hline
    Distributed Network & Inter-Planetary File System (IPFS) \cite{ipfs_doc} \\ \hline
    Message Queues & Firebase Cloud Firestore \cite{firestore} \\ \hline
    Content Pinning & Pinata \cite{pinata} \\ \hline
    User Authentication & Google Authentication and Firebase \cite{firebase} \\ \hline
    End-to-End Encryption & TweetNaCl.js \cite{tweetnacl}  \\ \hline
  \end{tabular}
\end{table}

We use WebRTC DataChannel for P2P message transmission when both the user devices are online. WebRTC DataChannel by default encrypts messages end-to-end and handles in-ordered messages. This is primarily useful when the user devices involved in the conversation are online, as it requires the users to have a persistent connection between each other. This reduces the delay for message transmission significantly.
In case of failure in establishment of DataChannel because of receiver being offline or otherwise, the message goes through the distributed network and for that we use Inter-Planetary File System (IPFS). IPFS is a public Distributed File Storage network available for various applications. For our experimentation, we need to have enough number of devices to run a distributed network and IPFS public network is able to deliver it. Message is pinned to other peers in the network using Pinata, it is a content pinning tool made for IPFS. It improves the availability of the message when the sender of the message is not connected to the internet by spreading chunks of the encrypted message to some IPFS nodes. Once the message is delivered to the receiver, the message is unpinned and eventually deleted from the distributed network.
The current implementation of message queues is done using Firebase by Google Cloud which is based on a centralized server architecture. The encrypted hashes of the undelivered messages are stored in a queue in a central server for bulk retrieval when the receiver is back online. The message queues will be later implemented in a distributed database which will ensure that even the hash of the encrypted messages are not stored with a central organization. This will be done using Inter-Planetary Naming System (IPNS), which is still in development. This will further reduce the cost of operating fybrrChat.
User Authentication in the current implementation is done using Google Authentication for easy and reliable authentication.
Finally for end-to-end encryption of the messages, Tweet-NaCl is used which uses public-key cryptography for encrypting the messages.


\begin{figure}[t]
     \centering
     \begin{subfigure}[b]{0.45\linewidth}
         \centering
         \frame{\includegraphics[width=\linewidth]{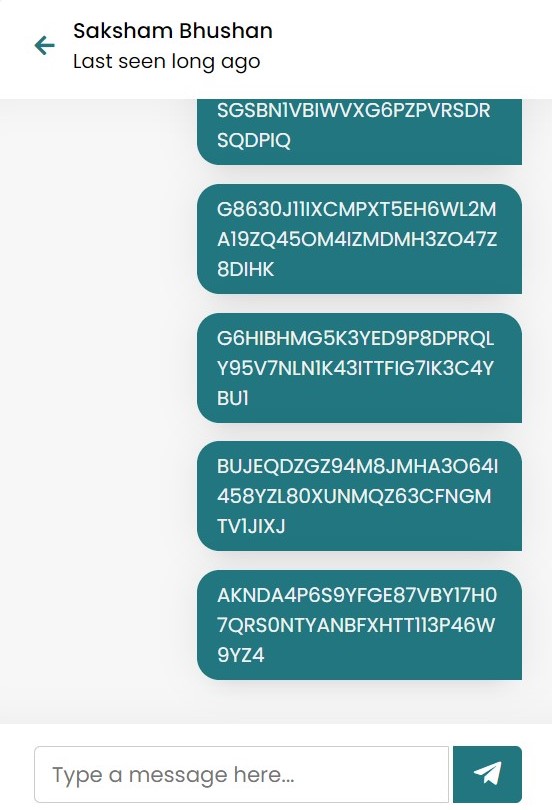}}
         \caption{Sending messages during experiment.}
         \label{fig:sssen}
     \end{subfigure}
     \hfill
     \begin{subfigure}[b]{0.45\linewidth}
         \centering
         \frame{\includegraphics[width=\linewidth]{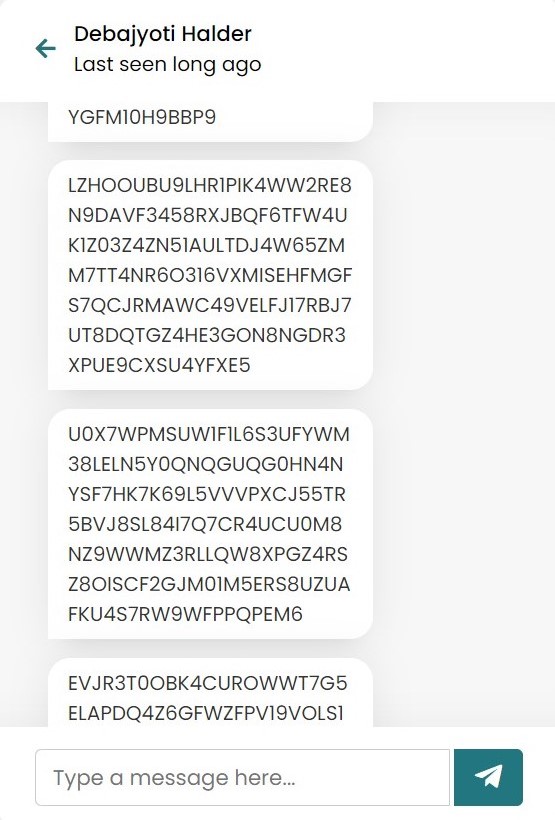}}
         \caption{Receiving messages during experiment.}
         \label{fig:ssrec}
     \end{subfigure}
        \caption{Screenshot of the application while performing the experiment mentioned in \ref{subsec:exp}.}
        \vspace{-1em}
        \label{fig:ss}
\end{figure}

\subsection{Experimental Results}\label{subsec:exp}
\begin{figure}
\centerline{\includegraphics[width=0.9\linewidth]{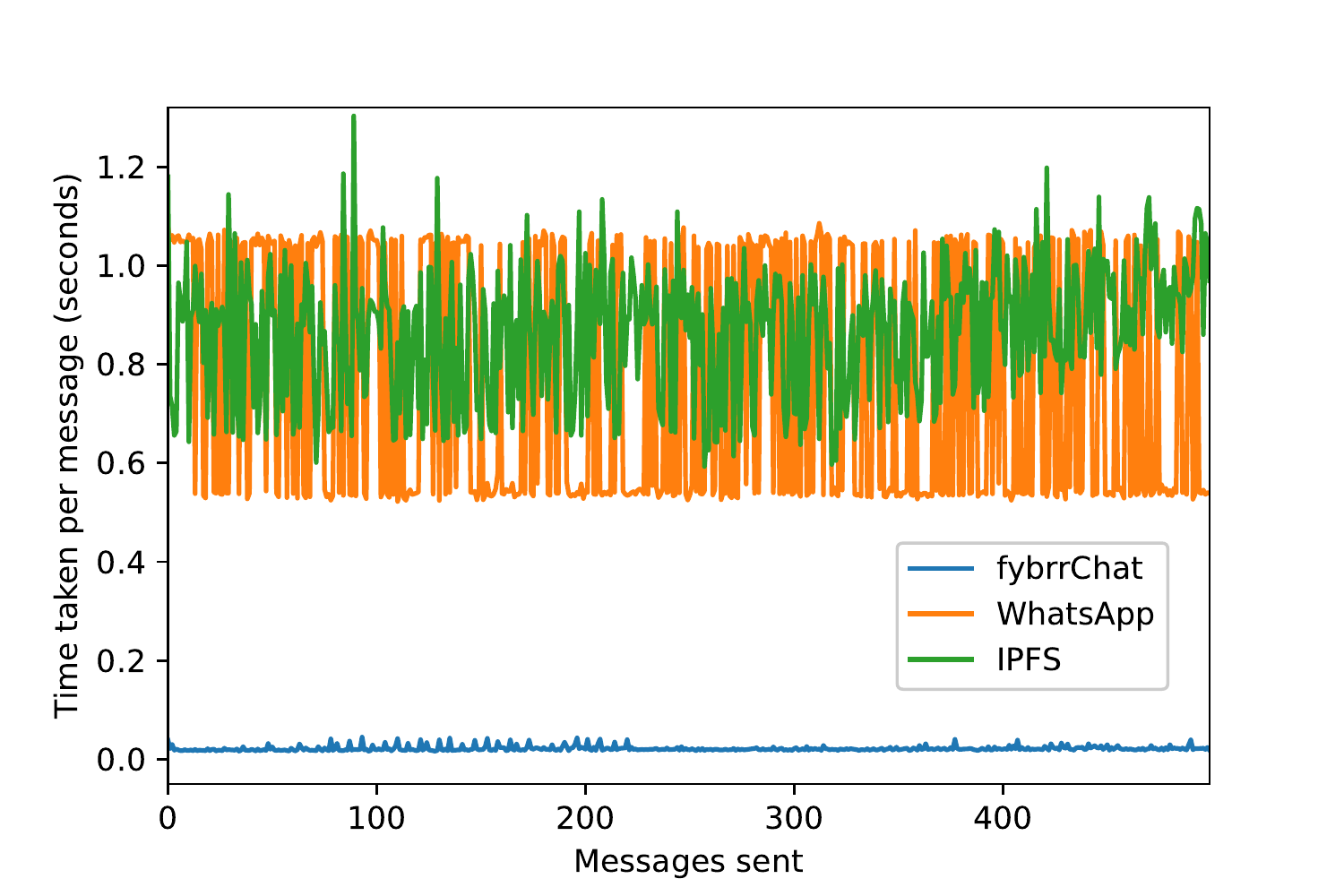}}
\caption{Time taken for sending each message using different schemes. fybrrChat represents the RTC DataChannel and IPFS represents the distributed network channel of fybrrChat.}
\vspace{-1em}
\label{fig:timetaken}
\end{figure}

We compared fybrrChat with WhatsApp in terms of time taken for receiving the messages. The simulation setup for WhatsApp was made using Selenium \cite{selenium}, which is a browser automation tool. The experiment for all schemes were done between the same two users, screenshots of the application window during the experiment can be seen in Fig. \ref{fig:ss}. A total of 500 messages containing random text of varying length were sent for each experiment. The length of messages varied linearly from 50 characters to 500 characters, excluding the metadata of the message. From Fig. \ref{fig:timetaken}, we can see that the RTC DataChannel delivers the message in significantly low time with an average of 0.022 seconds per message, whereas WhatApp takes an average of 0.771 seconds per message. Even with the distributed network channel (IPFS), fybrrChat is able to deliver the message in an average time of 0.866 seconds per message.

\begin{figure}
\centerline{\includegraphics[width=0.9\linewidth]{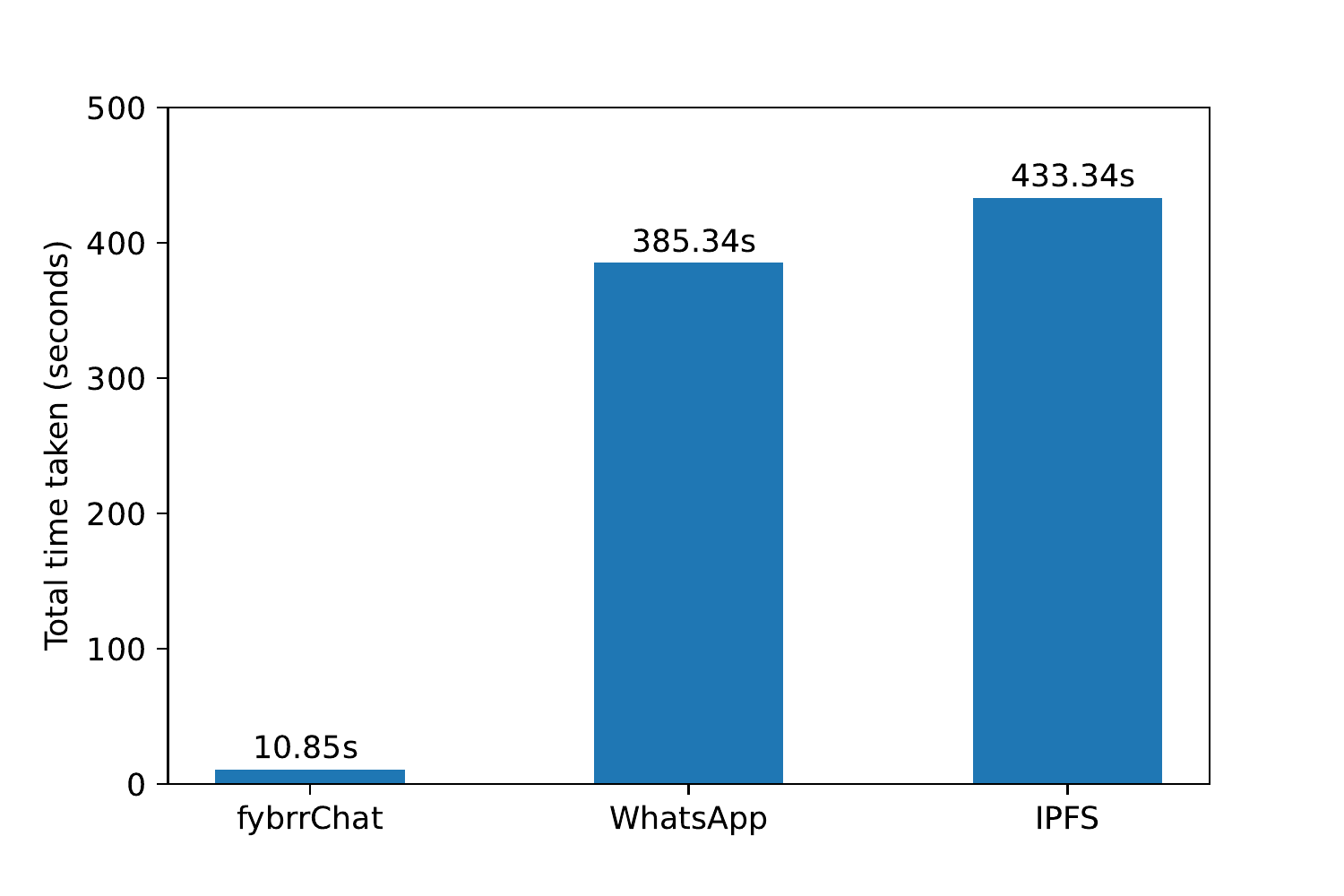}}
\caption{Total time taken to send 500 messages using different schemes.}
\vspace{-1em}
\label{fig:totaltime}
\end{figure}

Figure \ref{fig:totaltime} shows the total time taken to send 500 messages using different schemes. The RTC DataChannel is able to deliver all the messages within 11 seconds whereas WhatsApp takes around 385 seconds for delivering 500 messages, which shows that fybrrChat performs better in terms of time taken for sending messages. Although the IPFS channel takes 12.32\% more time than WhatsApp, it is used only when the receiver's device is not connected to the internet. Therefore, the receiver will receive the message within 1 second of connecting to the internet. For an active conversation, fybrrChat will use DataChannel, thus significantly reducing delay in message transmission. 

\subsection{Feasibility and Scalability}\label{subsec:feasibility}

To quantify the feasibility of a chat application assuming it to be $0.1\%$ of WhatsApp in terms of user count and messages sent. Using Azure Cloud servers, it will require almost 2000 Azure SignalR Servers which costs $\$0.671$ per hour per server, and $\$1$ per million messages. So the overall cost would amount to $\$100,966$ per month \cite{signalr}. These 2000 servers would consume atleast 14891.19 MWh of energy per year. This much energy can power atleast 59,500 suburban houses per year \cite{power}.

The peer-to-peer architecture of fybrrChat is much more scalable than WhatsApp and all other centralized messaging platforms due to its server-less architecture. The DHT in fybrrChat is also decentralized and uses the storage of devices using fybrrChat for intermediate message storage. The message queue data even if centralized consumes constant space per message (32 bytes) which makes it scalable. Even if we store 100 million message hashes in the queue per day, monthly costs for the Queue Storage servers of Azure required will amount to a maximum of $\$244$. This would consume 7.33 MWh of energy annually. The cost effectiveness of fybrrChat makes it feasible to scale up to multiple times than other text messaging applications. 


\section{Conclusion and Future Works}
The current social media platforms and instant messaging systems are always under the radar for lacking user privacy, exploitation of  users’ data and user censorship. We propose a decentralized secure trustless messaging architecture called fybrrChat which works without a central authority, thus ensuring that the users' data and privacy is not compromised. We compared our architecture with WhatsApp, by evaluating the time taken for messages to be delivered between two users. The experiment shows that fybrrChat is able to perform better than WhatsApp at even lesser cost of operation. The architecture can be extended from a messaging application to a full fledged social media platform, where the users are responsible for storage and processing of content within their devices instead of a central server as in fybrrChat.




\bibliographystyle{IEEEtran}
\bibliography{main}

\end{document}